\documentclass[12pt]{article}
\usepackage[dvips]{graphicx}
\usepackage{amssymb}
\begin{document}
\begin{center}
{\LARGE\bf Cosmologies with General Non-Canonical Scalar Field }
 \vskip 0.15 in
 $^\dag$Wei Fang$^1$, $^\ddag$H.Q.Lu$^1$, Z.G.Huang$^2$\\
$^1$Department~of~Physics,~Shanghai~University,\\
~Shanghai,~200444,~P.R.China\\
$^2$Department of Mathematics and Physics,
\\Huaihai Institute of Technology, Lianyungang, 222005, P.R.China
\footnotetext{$\dag$ xiaoweifang$_-$ren@shu.edu.cn}
\footnotetext{$\ddag$ Alberthq$_-$lu@staff.shu.edu.cn}

 \vskip 0.5 in
\centerline{\bf Abstract} \vskip 0.2 in
\begin{minipage}{5.5in} { \hspace*{15pt}\small We generally
investigate the scalar field model with the lagrangian
$L=F(X)-V(\phi)$, which we call it {\it General Non-Canonical Scalar
Field Model}. We find that it is a special square potential(with a
negative minimum) that drives the linear field
solution($\phi=\phi_0t$) while in K-essence model(with the
lagrangian $L=-V(\phi)F(X)$) the potential should be taken as an
inverse square form. Hence their cosmological evolution are totally
different. We further find that this linear field solutions are
highly degenerate, and their cosmological evolutions are actually
equivalent to the divergent model where its sound speed diverges. We
also study the stability of the linear field solution. With a simple
form of $F(X)=1-\sqrt{1-2X}$ we indicate that our model may be
considered as a unified model of dark matter and dark energy.
Finally we study the case when the baryotropic index $\gamma$ is
constant. It shows that, unlike the K-essence, the detailed form of
$F(X)$ depends on the potential $V(\phi)$. We analyze the stability
of this constant $\gamma_0$ solution and find that they are stable
for $\gamma_0\leq1$. Finally we simply consider the constant $c_s^2$
case and get an exact solution for $F(X)$.}
 { \hspace*{15pt}\small 
}
\end{minipage}
\end{center}
\newpage
\section{Introduction}
 \hspace{15pt}
Dark energy problem may be one of the biggest issues in current
theoretical physics and cosmology(see Ref[1] for a recent review).
The building of theoretical models as well as the breakthrough at
astrophysical observations have never halted since 1998. Technically
speaking, one can operate on either the r.h.s or the l.h.s of the
Einstein equation to get a reasonable interpretation of the
accelerating expansion of the universe. Many candidates of dark
energy have been proposed, such as the cosmological constant[2],
quintessence[3], K-essence[4], phantom[5], modifying gravity[6] and
so on. Among these models the scalar field models are undoubtedly
the most important class of theoretical models. Generally speaking,
the lagrangian of the scalar field model can be generally
represented as[7]
\begin{equation}L=f(\phi)F(X)-V(\phi)\end{equation}
where
$X=\frac{1}{2}\nabla_{\mu}\phi\nabla^{\mu}\phi=\frac{1}{2}{\dot{\phi}}^2
$ for a spatially homogeneous scalar field. Eq.(1) has included all
the popular single scalar field models. It describes K-essence when
$V(\phi)=0$ and standard quintessence when $f(\phi)=$constant and
$F(X)=X$. The idea of
 K-essence was firstly introduced as a possible model for inflation[8]
 and Later was considered as a possible model
 for dark energy[4,9]. L.P.Chimento found the first integral of the
 K-essence field equation for any function $F(X)$ when the potential
 is taken as inverse square form
or a constant[10]. In Ref[11], it was found that every quintessence
model can be view as a K-essence model generated by a kinetic linear
$F(X)$ function, and some K-essence potentials and their
quintessence correspondence are also found.
\par In this paper, we will focus on another class of models with
its lagrangian $L=F(X)-V(\phi)$, which we think is as important as
K-essence model but whose general characters and roles in cosmology
is far beyond clear. Here we should point out that, from the
original literature's point of view[8,9], our lagrangian(Eq.(2))
also belongs to K-essence model, which is charactered by a
lagrangian $L=L(X,\phi)$. However as far as we know, most works
about K-essence model are just based on the lagrangian form
$L=V(\phi)F(X)$. Therefore in this paper we call the model with
lagrangian $L=F(X)-V(\phi)$ {\it Non-Canonical Scalar Field Model}
while specify the model with the lagrangian $L=V(\phi)F(X)$ as
K-essence model. The paper is organized as follows: Section 2 is the
theoretical framework. In section 3 we investigate the linear field
solution and find the potential driving this linear field solution.
The divergent model with its speed of sound
diverge($c_s^2\rightarrow\infty$) is considered in section 4. Some
solvable general non-canonical scalar field solution are discussed
in section 5. Section 6 is the conclusion.
\section{Basic Framework}
\hspace{15pt} Let us restrict ourselves for the time being to the
cosmological setting corresponding to the flat universe described
by the FRW metric. We consider the spatially homogeneous real
scalar field $\phi$ with a non-canonical kinetic energy term. The
lagrangian density is
\begin{equation}L=F(X)-V(\phi)\end{equation}
where $V(\phi)$ is a potential and $F(X)$ is an arbitrary function
of $X$. Obviously above equation is a special case of Eq.(1) when
the function $f(\phi)=$constant. It includes quintessence [$F(X) =
X$] and a phantom field [$F(X) =-X$]. In fact this form of
lagrangian has appeared in Refs[1,12].
\par We can easily get the following equations:
\begin{equation}p=L=F(X)-V(\phi)\end{equation}
\begin{equation}\rho=3H^2=2L_{,X}X-L=2XF_{,X}-F(X)+V(\phi)\end{equation}
\begin{equation}{c_s}^2=p_{,X}/\rho_{,X}=[1+2X\frac{F_{,XX}}{F_{,X}}]^{-1}\end{equation}
Where we take $8\pi G=1$ for convenience. From Eqs.(3,4) we get the
relation $\rho+p=2XF_{,X}$ and this yields the state equation
$\omega_{\phi}$ larger than -1 if $F_{,X}>0$($\omega_{\phi}<-1$ if
$F_{,X}<0$). Vikman has argued that it is impossible for
$\omega_{\phi}$ to cross the phantom line divide($\omega_{\phi}=-1$)
in single scalar field theory[13]. However it is argued that this
result holds only for models without considering higher derivative
terms[14]. Eq.(5) describes the effective sound speed of the
perturbations. ${c_s}^2\equiv1$ if $F_{,XX}=0$ is satisfied.
\par The motion equations of the general non-linear scalar field
are
\begin{equation}\ddot\phi+3{c_s}^2H\dot\phi+\frac{\rho_{,\phi}}{\rho_{,X}}=0\end{equation}
\begin{equation}(F_{,X}+2XF_{,XX})\ddot\phi+3HF_{,X}\dot\phi+V_{,\phi}=0\end{equation}
\begin{equation}(\frac{\gamma}{\dot\phi})^{\cdot}+3H(1-\gamma)(\frac{\gamma}{\dot\phi})+\frac{V_{,\phi}}{3H^2}=0\end{equation}
where "$f_{,x}$" denotes the derivative with respect to subscript
index $x$ and $\gamma=(\rho+p)/\rho$ is the baryotropic index.
Eqs.(6,7,8) are different forms of the motion equation and they
are equivalent to each other.
\section{The Linear Field Model} \hspace*{15 pt}
In this section we will investigate a special case that the field
possesses a linear field solution $\phi=\phi_0t$. We find that the
form of potential $V(\phi)$ which drives this evolution is a square
potential with a negative minimum. We also show that the usual
linear field solution in non-linear scalar field theory leads to an
entirely different universe comparing with the universe in K-essence
model.

\par For the linear field solution, $X=\frac{1}{2}\dot\phi^2\equiv\frac{1}{2}{\phi_0}^2$ is a constant
and $\ddot\phi=0$, then we get following equation from Eqs.(4,7):
\begin{equation}\frac{{V_{,\phi}}^2}{3{F_{,X}}^2{\phi_0}^2}-V(\phi)+F(X)-{\phi_0}^2F_{,X}=0\end{equation}
$F$ and $F_{,X}$ are only the function of $X$ and therefore they
both are constant. We set $F_0=F(X_0)$ and $F_{,0}=F_{,X}(X_0)$ by
evaluating them at $X=X_0=\frac{1}{2}{\phi_0}^2$. Solving Eq.(9) and
Eq.(4) we get the exact solutions for potential $V(\phi)$ and scale
factor $a$:
\begin{equation}V(\phi)=\frac{3}{4}{F_{,0}}^2{\phi_0}^2(\phi+c)^2+F_0-F_{,0}{\phi_0}^2\end{equation}
\begin{equation}a=a_0exp[-\frac{F_{,0}}{4}(\phi_0t+c)^2]\end{equation}
Therefore the linear field solution leads to a square potential
Eq.(10) and a evolution of scale factor Eq.(11). It is worthwhile to
compare the same case in K-essence model. It is argued[15] that the
same linear field solution in K-essence model leads to an inverse
square potential and a power law expansion of scale factor. It is
interesting that the same linear field solution lead to different
cosmological evolution and therefore different cosmological
implication. We should emphasize that the potential is exactly
derived from Eqs.(4,9) and its form is unique. Moreover in this case
the different forms of $F(X)$ degenerate to only two cases:
$F_{,0}>0$ and $F_{,0}<0$ and respectively correspond to
non-phantom($\omega>-1$) and phantom case($\omega<-1$). The phantom
case($F_{,0}<0$) describes a universe from contracting phase to
expanding phase and is excluded easily by current observation. So we
restrict $F_{,0}>0$ for next discussion. From Eq.(4) we get
$\rho=F_{,0}{\phi_0}^2-F_0+V(\phi)$. To ensure the energy density
$\rho$ has a positive kinetic energy term we demand
$F_{,0}{\phi_0}^2-F_0>0$ and this immediately leads to the square
potential with a negative minimum value $F_0-F_{,0}{\phi_0}^2$[see
Eq.(10)]. In fact this result is well-intelligible in an expanding
universe. Because if the square potential has a non-negative minimum
it is well-known that the scalar field $\phi$ will roll down the
potential and finally cease at the minimum position $\phi=-c$ and
the linear field solution $\phi=\phi_0t$ will be no long valid. On
the other hand it is argued that for a potential with a negative
minimum the scalar field can oddly roll up the potential from its
minimum(see Fig.1) and the universe enters a contracting phase from
an expanding phase[16] and therefore the scalar field can evolve to
$\infty$.

\par It is very interesting that the universe in our model can avoid
a beginning singularity. If we think the classic cosmology is valid
when energy scale is below Plank scale, then the scale factor at the
beginning is very small[$a=a_0e^{-\rho_{pl}/(3F_{,0}{\phi_0}^2)}$,
where $\rho_{pl}$ is the energy density at plank time] but does not
equal zero. However this universe can not escape from a collapse in
future. This evolutive behaviors are completely different from the
same linear field solution case($\phi=\phi_0t$) in K-essence model
where the scale factor behaves as $a\propto t^n$[10] and the
universe was birth from a singularity and expand for ever.

\par One of our concerns is whether our model can describe a suitable
universe with a phase of accelerating expansion. The answer is
positive because we have $\ddot
a\propto-(\rho+3p)=-\frac{3}{2}F_{,0}{\phi_0}^2[2-F_{,0}(\phi+c)^2]$
and $\rho+3p<0$ for  $\phi<\phi_1$ or $\phi>\phi_2$, where
$\phi_1=-c-\sqrt{\frac{2}{F_{,0}}},
\phi_2=-c+\sqrt{\frac{2}{F_{,0}}}$. The potential and the evolutive
behaviors of universe are showed in Fig.1. We can see from Fig.1,
the field rolls down the potential from an initial value and the
universe undergoes an accelerating expansion. When the field
evolutes to $\phi_1$, the universe enters a decelerating expansion
and finally becomes zero expansion rate when the field arrives at
$\phi=-c$. When the field crosses the point $\phi=-c$ the expansion
rate $H$ dramatically changes it sign from $H>0$ to $H<0$ and the
field rolls up the potential from its minimum. When the field rolls
from $-c$ to $\phi_2$ the universe undergoes an accelerating
contraction. After going over $\phi_2$, the universe enters a
decelerating contraction and collapses to singularity finally.
  \vskip 0.3
in
\begin{center}
\includegraphics[scale=1,origin=c,angle=0]{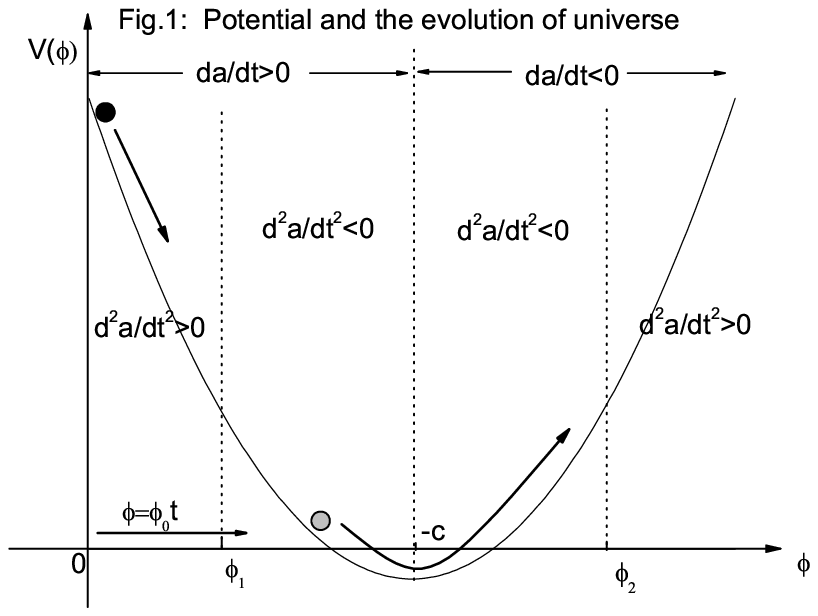}
\end{center}
\par In
addition, since we get the potential(Eq.(10)) just from the linear
field solution without any other assumption, another non-trivial
question is whether an arbitrary square potential with a negative
constant can always lead to the linear field solution.
Unfortunately, we will demonstrate that it is not the case: not all
the square potential with a negative constant can possess the linear
field solution.

\par Given a arbitrary potential:
\begin{equation} V(\phi)=A(\phi+c)^2-B\end{equation}
where A, B are arbitrary positive constants. In order to have a
linear field solution, from Eq.(10) A and B should satisfy
$A=\frac{3}{4}{F_{,0}}^2{\phi_0}^2, B=F_0-F_{,0}{\phi_0}^2$. So we
get the following constraint equation:
\begin{equation} 4A=3F_{,0}(F_0-B)\end{equation}
It means that A and B are not arbitrary parameters. In other words,
given a value of A, the value of B is determined by Eq.(13). i.e,
only the square potential with its parameters satisfying Eq.(13) can
lead to a linear field solution. From Eq.(13) we find another
interesting character for the linear field solution. Since there
only appears the first two coefficients ($F_0, F_{,0}$) of the
series expansion of the function $F(X)$ around $X=X_0$(we can expand
function $F(X)$ as $F_0+F_{,0}(X-X_0)+\cdots$ around $X_0$). The
evolution of universe will be thought of as equivalent as long as
the function $F(X)$ has the same first two coefficients of its
expansion, disregarding the rest of the higher order terms. This
means that the linear filed solution model possesses a high
degenerate character.
\par The last question we ask is, how stable
are the linear field solutions. Since for the linear field solution
we have $3HF_{,0}\phi_0+V_{,\phi}=0$, we can rewrite Eq.(7) as
follows:
\begin{equation}\frac{d\dot\phi}{dt}=\frac{-3H}{F_{,X}+2XF_{,XX}}(F_{,X}\dot\phi-F_{,0}\phi_0)\end{equation}
For the case that $c_s^2=costant\geq0$, Eq.(14) becomes
\begin{equation}\frac{d\dot\phi}{dt}=-3Hc_s^2(\dot\phi-\phi_0)\end{equation}
Integrating Eq.(15), we have
\begin{equation}\dot\phi=\phi_0+\frac{c_1}{a^{3c_s^2}}\end{equation}
Eq.(16) shows that, if the universe expands for ever, $\dot\phi$ has
an asymptotic limit $\phi_0$ and the solution $\phi=\phi_0 t$ is
stable. However, as we known from Eq.(11), after expanding to a
maximum scale factor the universe will contract to a singularity
eventually. Therefore this linear field solution can not be stable.
\section{The Divergent Model}
In this section we investigate another interesting model where the
speed of sound is divergent($c_s^2\rightarrow\infty$). From Eq.(5),
we have:
\begin{equation}1+2X\frac{F_{,XX}}{F_{,X}}=0\end{equation}
Integrating Eq.(17) we get the form of function $F(X)$:
\begin{equation}F(X)=c_2X^{\frac{1}{2}}+c_3\end{equation}
where $c_2, c_3$ are the integral constants. From Eq.(18), the sound
 speed of this special form of $F(X)$ diverges(${c_s}^2\rightarrow\infty$). The same form of
function $F(X)$ is also obtained in K-essence model with the
lagrangian being $-V(\phi)F(X)$[10,15]. Recently This type
lagrangian is thoroughly investigated and exploited in model
building[17], where this type lagrangian is called as {\it
Cuscuton}. From Eq.(4) and Eq.(18), we have
\begin{equation}3H^2=V(\phi)-c_3\end{equation}
From Eq.(7) and Eq.(19) we get
\begin{equation}\frac{1}{2}c_2\sqrt{3(V(\phi)-c_3)}+dV(\phi)/d\phi=0\end{equation}
Therefore we can immediately get the potential $V(\phi)$:
\begin{equation}V(\phi)=\frac{3}{8}c_2^2(\phi-c_4)^2+c_3\end{equation}
where $c_4$ is an integral constant. For the model building the
concrete form of function $F(X)$ and the potential $V(\phi)$ can be
constructed respectively, however, it is very interesting that the
divergent model with the special lagrangian Eq.(18) determine the
unique form of potential Eq.(21). This maybe means that the
divergent model with the speed of sound $c_s^2=\infty$ has some
special implications. Let us recall the result obtained in Section
3: if the square potential satisfies the constraint Eq.(10), the
solution of the scalar field will be linear field solution. From
Eqs.(10, 18), we have
$\frac{3}{4}{F_{,0}}^2{\phi_0}^2=\frac{3}{8}c_2^2,
F_0-F_{,0}{\phi_0}^2=c_3$, which just coincides with Eq.(21). This
means that the divergent model and the linear field solution are
degenerate. Namely the divergent model and the linear field solution
model are kinetically isomorphic and share the same evolution of
scale factor $a$ and scalar field $\phi$:
\begin{equation}\phi=\phi_0t,
~~~a=a_0exp[-\frac{F_{,0}}{4}(\phi_0t+c)^2]\end{equation} However it
remains to clarify why this could happen. From the mathematical
point of view, we maybe get the interpretation from the motion
equation of field Eq.(7) since both the divergent model and the
linear field model lead the first term of Eq.(7) to vanish.
Therefore they share the same motion equation and then lead to the
same cosmological evolution. But what is the physical implication
that the scalar field theory with an infinite speed of sound is
degenerate with the linear field model? There also exists the same
situation in the K-essence model that the linear field solution and
the divergent model($c_s^2=\infty$) are isomorphic[18].
\section{Solvable General Non-canonical Scalar Field Cosmologies}
\hspace{15pt} In this section we will focus on some special cases
when Eq.(8) exists a first integral or can be solved exactly. Though
what we consider are quite simple, they can also lead to some
important results. Additional, we will try to find the relationships
and the differences between our General Non-Canonical Scalar Field
model and K-essence model.
\par \textbf{A. $V(\phi)=V_0$}
\par When the potential $V(\phi)$ is a constant($=V_0$),  Eq.(8)
exists a first integral
\begin{equation}\frac{\gamma}{\dot\phi}=\frac{c_5}{a^3H^2}\end{equation}
where $c_5$ is an arbitrary integral constant. Eq.(23) is very
similar with the first integral obtained in K-essence model with a
constant potential[10,19]. For a constant potential the K-essence
lagrangian can be written as $L_k=-V_0F_k(X)$ while the lagrangian
 in our model is $L_g=F_g(X)-V_0$. They are actually equivalent if we define
$F_g(X)=V_0(1-F_g(X))$. So our model can easily reproduce the
K-essence models with constant potential. The K-essence models with
a constant potential are hotly studied for its exquisite role in
unifying the dark matter and dark energy[20]. Let we consider a
special form of lagrangian $L_g=1-\sqrt{1-2X}-V(\phi)$, which is
considered as a Non-Linear Born-Infeld(NLBI) scalar field theory in
Ref[21]. When the potential is constant $V_0$, we can find the exact
solutions:
\begin{equation}{\dot\phi}^2=\frac{c_6}{c_6+a^6}\end{equation}
\begin{equation}\rho=\sqrt{1+\frac{c_6}{a^6}}+(V_0-1)\end{equation}
\begin{equation}{c_s}^2=1-{\dot\phi}^2=\frac{a^6}{c_6+a^6}\end{equation}
Where $c_6$ is an integral constant. From Eqs.(25,26) we know that,
the energy density behaves as dark matter at early time(for small
$a$, $\rho\propto a^{-3}$ and ${c_s}^2\simeq 0$) and dark energy at
late time(for large $a$, $\rho\simeq V_0=const$ and ${c_s}^2\simeq 1
$). So, our model can also play the same role that unifies the dark
matter and dark energy.

\par \textbf{B. $\gamma=$constant}
\par In this subsection we will assume that the baryotropic index
$\gamma=\gamma_0=$constant. Then the state equation
$\omega=\gamma-1=\gamma_0-1$, is also a constant. The constant
$\gamma$ kinematically leads to the cosmological solution
\begin{equation}a=a_0t^{2/3\gamma_0},~~~\rho_{\phi}=\frac{4a_0^{3\gamma_0}}{3\gamma_0^2}/a^{3\gamma_0}\end{equation}
 From the relation
 \begin{equation}\gamma_0=\frac{\rho+p}{\rho}=-\frac{2\dot H}{3H^2}=\frac{2XF_{,X}}{2XF_{,X}-F(X)+V(\phi)}\end{equation}
 We get the following equation:
\begin{equation}\frac{2(\gamma_0-1)}{\gamma_0}XF_{,X}-F(X)+V(\phi)=0\end{equation}
In Ref[18], it is showed that
$F(X)=X^{\frac{\gamma_0}{2(\gamma_0-1)}}$ can lead to the constant
$\gamma$ with any potential in K-essence model. Here we show that in
our model the form of $F(X)$ depends on the potential(see Eq.(29)).
Namely, to admit the cosmological solution Eq.(27), the kinetic term
$F(X)$ and potential term $V(\phi)$ must satisfy Eq.(29). Only for a
constant potential $V_0$ ,we get a similar function of $F(X)$:
\begin{equation}F(X)=c_7X^{\frac{\gamma_0}{2(\gamma_0-1)}}+V_0\end{equation}
where $c_7$ is an integral constant.
\par It is quite interesting to consider whether the solution with constant
$\gamma_0$ is stable. To answer this question, we let $\gamma$ vary
with time. Differentiating the equation of the baryotropic index
$\gamma$, we get
\begin{equation}\dot\gamma=(\gamma-1)(3H\gamma+\frac{\dot p}{p})\end{equation}
We immediately get two critical points: $\gamma_0-1=0$ or $\gamma_0$
satisfies Eq.(32):
\begin{equation}3H\gamma_0+\frac{\dot p}{p}=0\end{equation}
When this stationary condition Eq.(32) holds, the potential
$V(\phi)$ and function $F(X)$ will satisfy the relation:
\begin{equation}p=F(X)-V(\phi)=\frac{c_8}{a^{3\gamma_0}}\end{equation}
From Eqs.(31,32), we have
\begin{equation}\dot\gamma=3H(\gamma-1)(\gamma-\gamma_0)\end{equation}
Integrating Eq(34) we get
\begin{equation}\gamma=\frac{\gamma_0a^{3(1-\gamma_0)}-c_9}{a^{3(1-\gamma_0)}-c_9}\end{equation}
where $c_8,~c_9$ is an integral constant. Eq.(35) indicates that,
for the expanding universe and $\gamma_0<1$ the baryotropic index
$\gamma$ has the asymptotic limit $\gamma_0$. However for
$\gamma_0>1$ the baryotropic index $\gamma$ will approach the
asymptotic limit $1$. The case $\gamma_0=1$ should be considered
apart. For $\gamma_0=1$, the solution is
\begin{equation}\gamma=1-\frac{c_{10}}{3ln a}\end{equation}
Where $c_{10}$ is an integral constant. Eq.(36) shows that the
solution with $\gamma_0=1$ is also stable in an expanding universe.
Therefore, we can conclude that the solutions with constant
baryotropic index are attractors in the case $\gamma_0\leq1$ and the
$\gamma_0=1$ solutions separate stable from unstable regions in the
phase space.
\par \textbf{C. $c_s^2=$constant}
\par For the dark energy behaving as a fluid, the speed of sound $c_s$ is another important parameter
 in addition to the equation of state $\omega$. The speed of sound
$c_s$ is the propagation of the perturbation of the background
scalar field, which can affects the CMB power spectrum. Therefore
the effective sound speed $c_s$ of dark energy would provide crucial
information which is complementary to the equation of state
$\omega$. In this subsection we will study the simple case that the
speed of sound is constant. From Eq.(5), we obtain the equation as
follows:
\begin{equation}2c_s^2XF_{,XX}=(1-c_s^2)F_{,X}\end{equation}
Integrating Eq.(26), we have
\begin{equation}F(X)=\frac{2c_s^2}{1+c_s^2}c_{11}X^{\frac{1+c_s^2}{2c_s^2}}+c_{12}\end{equation}
where $c_{11},~c_{12}$ is the integral constants.
\section{Conclusion}
\hspace*{15 pt}In this paper we have generally investigated the
general non-canonical scalar field model as a candidate of dark
energy. We found that it was a special square potential(with a
negative minimum) that drove the linear field solution in our model
while the potential should be taken as an inverse square form in
K-essence model. Our results showed that the linear field solution
was highly degenerate, and shared the same cosmological evolution
with the divergent model where its sound speed
$c_s^2\rightarrow\infty$. We pointed out that our model with a
constant potential was actually equivalent to the K-essence model
also with a constant potential. With a simple form of $F(X)$ we
indicated that our model can be also considered as a unified model
of dark matter and dark energy. In addition we studied the constant
$\gamma_0$ case. The results showed that, unlike the K-essence
model, the detailed form of $F(X)$ depended on the potential. We
find that the constant $\gamma_0$ solution is stable for
$\gamma_0\leq1$. We also found the form of $F(X)$ which possessed
the constant $c_s^2$ solution. Our work may throw light on the study
of the scalar field theory and the exploration of dark energy.
\section{Acknowledgement}
 \hspace*{15 pt}W.Fang would like to thank Dr.Hongwang Yu and Xingyu Jin for useful discussion.
 We also appreciate the anonymous referee for the helpful suggestion.
 This work is partly supported by NNSFC under Grant No.10573012 and No.10575068 and by Shanghai
 Municipal Science and Technology Commission No.04dz05905. \\

{\noindent\Large \bf References} \small{
\begin{description}
\item {1.} {E.J.Copeland, M.Sami and S.Tsujikawa, hep-th/0603057}
\item {2.} {S.Weinberg, Rev.Mod.Phys\textbf{61}(1), 1(1989);\\
             P.J.Steinhardt, L.Wang and I.Zlatev,Phys.Rev.D\textbf{59}, 123504(1999);\\
             P.J.E.Peebles and B.Ratra, Rev.Mod.Phys\textbf{75}(2), 559(2003);\\
             A.G.Riess et al., Astrophys.J\textbf{607}, 665-687(2004);\\
             G.W.Gibbons, hep-th/0302199.}
\item {3.} {R.R.Caldwell, R.Dave and P.J.Steinhardt, Phys.Rev.Lett.\textbf{80}, 1582(1998);\\
             P,J.Steinhardt,L.Wang and I.Zlatev, Phys.Rev.Lett.\textbf{82}, 896(1996);\\
             X.Z.Li, J.G.Hao and D.J.Liu, Class.Quantum Grav.\textbf{19}, 6049(2002).}
\item {4.} {J.M.Aguirregabiria, L.P.Chiemento and R.Lazkoz, Phys.Rev.D\textbf{70}, 023509(2004);\\
             E.J.Copeland, M.R.Garousi, M.Sami and S.Tsujikawa, Phys.Rev.D\textbf{71}, 043003(2005);\\
             G.Panotopoulos; astro-ph/0606249;\\
             M.Malquarti, E.J.Copeland and A.R.Liddle, Phys.Rev.D\textbf{68}, 023512(2003);\\
             R.Lazkoz, Int.J.Mod.Phys.D\textbf{14}, 635-642 (2005).}
\item {5.} {R.R.Caldwell, Phys.Rev.Lett.B\textbf{545}, 23(2002);\\
             M.Sami, Mod.Phys.Lett.A\textbf{19}, 1509(2004);\\
             M.Sami and T.Padamanabhan, Phys.Rev.D\textbf{67}, 083509(2003);\\
             M.Sami, P.Chingangbam and T.Qureshi, hep-th/0301140;\\
             P.Singh, M.Sami, N.Dadhich,Phys.Rev.D\textbf{68}, 023522(2003);\\
             V.Faraoni, Class.Quant.Grav.\textbf{22}, 3235-3246(2005);\\
             D.J.Liu and X.Z.Li, Phys.Rev.D\textbf{68}, 067301(2003);\\
             M.P.Dabrowski et.al., Phys.Rev.D\textbf{68}, 103519(2003);\\
             S.M.Carroll, M.Hoffman and M.Teodden, Phys.Rev.D\textbf{68}, 023509(2003);\\
             Y.S.Piao, R.G.Cai, X.M.Zhang and Y.Z.Zhang, Phys.Rev.D\textbf{66}, 121301(2002);\\
             S.Mukohyama, Phys.Rev.D\textbf{66}, 024009(2002);\\
             T.Padmanabhan, Phys.Rev.D\textbf{66}, 021301(2002);\\
             G.Shiu and I.Wasserman, Phys.Lett.B\textbf{541}, 6(2002);\\
             L.kofman and A.Linde, JHEP\textbf{0207}, 004(2002);\\
             A.Sen, JHEP\textbf{0210}, 003(2002);\\
             I.Ya.Aref'eva and A.S.Koshelev, hep-th/0605085;\\
             I.Ya.Aref'eva, A.S.Koshelev and S.Yu.Vernov, Phys.Rev.D\textbf{72}, 064017(2005);\\
             I.Ya.Aref'eva, A.S.Koshelev and S.Yu.Vernov, Phys.Lett.B\textbf{628}, 1-10(2005);\\
             I.Ya.Aref'eva, A.S.Koshelev and S.Yu.Vernov, astro-ph/0412619;\\
             N.Moeller and B.Zwiebach, JHEP\textbf{0210}, 034(2002);\\
             P.Mukhopadhyay and A.Sen, JHEP\textbf{0211}, 047(2002);\\
             T.Okunda and S.Sugimoto, Nucl.Phys.B\textbf{647}, 101-116(2002);\\
             G.Gibbons, K.Hashimoto and P.Yi, hep-th/0209034;\\
             B.Chen, M.Li and F.Lin, hep-th/0209222;\\
             G.Felder, L.Kofman and A.Starobinsky, JHEP\textbf{0209}, 026(2002);\\
             M.C.Bento, O.Bertolami and A.A.Sen, hep-th/0208124;\\
             H.Lee et.al., hep-th/0210221;\\
             J.G.Hao and X.Z.Li, Phys.Rev.D\textbf{66}, 087301(2002); Phys.Rev.D\textbf{68}, 043501(2003);\\
             X.Z.Li and X.H.Zhai, Phys.Rev.D\textbf{67}, 067501(2003);\\
             Z.G.Huang, H.Q.Lu and W.Fang, Class.Quant.Grav\textbf{23}, 6215(2006);\\
             S.Nojiri and S.D.Odintsov, Phys.Lett.B\textbf{562}, 147(2003);\\
             E.Elizalde, S.Nojiri and S.D.Odintsov, Phys.Rev.D\textbf{70}, 043539(2004);\\
             S.Nojiri and S.D.Odintsov, Phys.Rev.D\textbf{70}, 103522(2004);\\
             S.Nojiri, S.D.Odintsov and S.Tsujikawa, Phys.Rev.D\textbf{71},063004(2005);\\
             S.Nojiri and S.D.Odintsov, Phys.Rev.D\textbf{72}, 023003(2005).}
\item {6.} {G.Dvali, G.Gabadadze and M.Porrati, Phys.Lett.B\textbf{485}, 208(2000);\\
             G.Dvali, G.Gabadadze, M.Kolanovic and F.Nitti, Phys.Rev.D\textbf{64}, 084004(2001);\\
             G.Dvali, G.Gabadadze, M.Kolanovic and F.Nitti, Phys.Rev.D\textbf{65}, 024031(2002);\\
             C.Deffayet, Phys.Lett.B\textbf{502}, 199(2001);\\
             C.Deffayet, G.R.Dvali and G.Gabadadze, Phys.Rev.D\textbf{65}, 044023(2002);\\
             S.Nojiri and S.D.Odintsov, hep-th/0601213;\\
             B.Li and M.C.Chu, Phys.Rev.D\textbf{74}, 104010(2006);\\
             B.Li, K.C.Chan and M.C.Chu, astro-ph/0610794.}
\item {7.} {A.Melchiorri, L.Mersini, C.J.Odman and M.Trodden, Phys.Rev.D\textbf{68}, 043509(2003).}
\item {8.} {J.Garriga and V.F.Mukhanov, Phys.Lett.B\textbf{458}, 219(1999);\\
             C.Armend\'{a}riz-Pic\'{o}n, T.Damour and V.Mukhanov, Phys.Lett.B\textbf{458}, 209(1999).}
\item {9.} {C.Armend\'{a}riz-Pic\'{o}n, V.Mukhanov and P.J.Steinhardt, Phys.Rev.Lett\textbf{85}, 4438(2000);\\
             C.Armend\'{a}riz-Pic\'{o}n, V.Mukhanov and P.J.Steinhardt, Phys.Rev.D\textbf{63}, 103510(2001);\\
             T.Chiba, Phys.Rev.D\textbf{66}, 063514;\\
             T.Chiba, T.Okabe and M.Yamaguchi, Phys.Rev.D\textbf{62}, 023511(2000);\\
             M.Malquarti, E.J.Copeland, A.R.Liddle and M.Trodden, Phys.Rev.D\textbf{67}, 123503(2003).}
\item {10.} {L.P.Chimento, Phys.Rev.D\textbf{69}, 123517(2004). }
\item {11.} {J.M.Aguirregabiria, L.P.Chimento and R.Lazkoz, Phys.Lett.B\textbf{631}, 93-99(2005). }
\item {12.} {V.Mukhanov and A.Vikman, JCAP\textbf{0602}004 (2006); A.Vikman, astro-ph/0606033\\
             E.Babichev, hep-th/0608071.}
\item {13.} {A.Vikman, Phys.Rev.D\textbf{71}, 023515(2005).}
\item {14.} {M.Li, B.Feng and X.Zhang, JCAP\textbf{0512},002(2005);\\
             A.Anisimov, E.Babichev and A.Vikman, JCAP\textbf{0506},006(2005);\\
             X.F.Zhang and T.Qiu, astro-ph/0603824.}
\item {15.} {L.P.Chimento and A.Feinstein, Mod.Phys.Lett.A\textbf{19}, 761-768 (2004).}
\item {16.} {G.Felder, A.Frolov, L.Kofman and A.Linde, Phys.Rev.D\textbf{42}, 023507(2002);\\
             A.de la Macorra and C.Stephan, Phys.Rev. D\textbf{65}, 083520(2002);\\
             Imogen P. C. Heard and David Wands, Class.Quant.Grav.\textbf{19}, 5435-5448(2002);\\
             L.Perivolaropoulos, Phys.Rev. D\textbf{71}, 063503(2005);\\
             A.de la Macorra and G. German, astro-ph/0212148;\\
             W.Fang, H.Q.Lu and Z.G.Huang, hep-th/0606033.}
\item {17.} {N.Afshordi, D.J.H.Chung and G.Geshnizjani, hep-th/0609150.}
\item {18.} {L.P.Chimento and A.Feinstein, Mod.Phys.Lett.A\textbf{19}, 761-768 (2004).}
\item {19.} {C.Armend\'{a}riz-Pic\'{o}n,T.Damour and V.Mukhanov,Phys.Lett.B\textbf{458}, 209(1999).}
\item {20.} {R.J.Scherrer, Phys.Rev.Lett\textbf{93}, 011301(2004);\\
              L.P.Chimento, M.Forte and R.Lazkoz, Mod.Phys.Lett.A\textbf{20}, 2075(2005);\\
              M.Novello, M.Makler, L.S.Werneck and C.A.Romero, Phys.Rev.D\textbf{71}, 043515(2005)}
\item {21.} {H.Q.Lu, Int.J.Mod.Phys.D\textbf{14}, 355(2005);\\
             W.Fang, H.Q.Lu, Z.G.Huang and K.F.Zhang, Int.J.Mod.Phys.D\textbf{15}, 199-214(2006)(hep-th/0409080);\\
             W.Fang, H.Q.Lu and B.Li and K.F.Zhang, Int.J.Mod.Phys.D\textbf{15}, 1947-1961(2006)(hep-th/0512120);\\
             Z.G.Huang, X.H.Li and Q.Q.Sun, hep-th/0610019 }

\end{description}
\end{document}